\documentclass[twoside,showpacs,showkeys,
  preprintnumbers,superscriptaddress,amsmath,amssymb,pra]{revtex4}
\usepackage[dvips]{graphicx, color}
\usepackage{verbatim}
\newcommand{\ket}[1]{\ensuremath{|#1\rangle}}
\newcommand{\bra}[1]{\ensuremath{\langle#1|}}

\newcommand{\degree}{\ensuremath{^\circ}}

\begin{document}
\title{Demonstration of active routing of entanglement in a multi-user network}

\newcommand{\UniWien}{Quantum Optics, Quantum Nanophysics and Quantum Information, Faculty of Physics, University of Vienna, Boltzmanngasse 5, 1090 Vienna, Austria}
\newcommand{\AcaWien}{Institute for Quantum Optics and Quantum Information,
Austrian Academy of Sciences, Boltzmanngasse 3, 1090 Vienna,
Austria}  
\newcommand{\Stock} {Department of Physics, Stockholm University, 109 61 Stockholm, Sweden}
\newcommand{\IQC}{Institute for Quantum Computing and Department of
Physics \& Astronomy, University of Waterloo, Waterloo N2L 3G1, Canada}
\newcommand{\ARC}{Optical Quantum Technologies, Safety \& Security Department, AIT Austrian Institute of Technology,
  Donau-City-Stra\ss e~1, 1220 Vienna, Austria}

\author{I~Herbauts}
\email[Corresponding author:~]{isabelle.herbauts@fysik.su.se}
\affiliation{\Stock}
\affiliation{\UniWien}

\author{B~Blauensteiner}
\affiliation{\UniWien}

\author{A~Poppe}
\affiliation{\ARC}

\author{T~Jennewein} 
\affiliation{\IQC}
\affiliation{\AcaWien}
\author{H~H\"{u}bel}
\affiliation{\Stock}
\affiliation{\UniWien}

\date{\today}

\begin{abstract}
We implement an entanglement distribution network based on wavelength-multiplexing and optical switching for quantum communication applications.
Using a high-brightness source based on spontaneous parametric
down-conversion in periodically-poled lithium niobate waveguides, we generate polarisation entangled photon pairs with a broad spectrum covering the telecom wavelengths around 1550~nm.
The photon pairs have entanglement fidelities up to $99\%$, and are distributed via passive wavelength multiplexing in a static multi-user network. We furthermore demonstrate a possible network application in a
scenario with a single centralised source dynamically allocating two-party entanglement to any pair of users by means of optical switches.
The whole system, from the pump laser up to the receivers, is
fibre and waveguide based, resulting in maximal stability, minimal
losses and the advantage of readily integrable telecom components in
the $1550$~nm range.

\end{abstract}


\preprint{arXiv:}

\maketitle


\section{Introduction}

Quantum entanglement is an essential resource for many quantum information experiments, such as quantum teleportation, entanglement swapping and tests of Bell inequalities~\cite{pan12}. It is also useful in quantum communication protocols such as Quantum Key Distribution (QKD)~\cite{zbi02}. For QKD in particular, entanglement offers the additional advantage of device independent security~\cite{sca07}.    Entanglement-based QKD is now a well developed research area which has, in the last years, reached important practical milestones~\cite{yam08, hub09, sch09, kur09}. Forefront developments in the field include also the realisation of more sophisticated multi-party QKD networks~\cite{secoq,tokyo}, with trusted nodes connecting different independent QKD-links. So far however, only photonic qubits were distributed in these demonstrations, but no entanglement. The realisation of \emph{entanglement-distribution} networks connecting multiple users sharing entanglement would permit a wide variety of quantum communication applications in addition to QKD such as secret key sharing~\cite{ber99} and quantum complexity protocols~\cite{dam01}.

Entangled photon sources (EPS) used in  such networks should incorporate design features to integrate them easily into the existing telecommunication infrastructure. It is therefore advantageous to have compact, cheap and power-saving systems which are also compatible with fibre-based telecom components. For practical ground-based networks, the generation of entangled photons around 1550~nm is beneficial, since long distance transmission in fibre is optimal at that wavelength and there exist standardised grids in the form of Dense Wavelength Division Multiplexing (DWDM) and Coarse Wavelength Division Multiplexing (CWDM)~\cite{itu}.

A common technique to create and make practical use of entanglement is to produce polarization-entangled photons by spontaneous parametric down-conversion (SPDC) in second order ($\chi^2$) non-linear crystals~\cite{kwiat1995nhi}. The coupling of photons from the bulk crystal to optical fibres leads to a loss of robustness and reliability, both needed in practical applications. Non-linear crystals with inscribed waveguides and fibre pigtails overcome this problem and offer long term stability. In addition, waveguides also increase the efficiency of the down-conversion process by several orders of magnitude compared to bulk crystals~\cite{gis01}. 
Entangled photon pairs have also been directly generated in dispersion shifted fibres via four-wave mixing techniques ($\chi^3$)~\cite{LCLLVK06}. Such designs offer the advantage of an all-fibre based compact system with no coupling losses from crystal to fibre. The $\chi^3$ process however suffers from Raman scattering, leading to an increase in background photons. To limit the scattering, fibres have to be cooled  with liquid nitrogen or the produced photon pair has to lie outside the scattering band ($\Delta \lambda \sim 250$~nm)~\cite{kan09,gov07,jen12}. Neither option is very practical for realistic telecom applications since cooling requires more maintenance and larger source designs, or the photons would lie outside the telecommunication band (1260-1675~nm), due to the large wavelength separation.  Recent developments include the generation of entangled photon pairs in a twin-hole step-index fibre via $\chi^2$ nonlinearities~\cite{li12}. Although this seems a  promising way for the future, conversion efficiencies of $\chi^2$-fibres are currently still much lower than in $\chi^2$-crystals with waveguides which are the best candidates for  practical realisations.

\begin{figure}[t]
\begin{center}
\includegraphics[width=8cm,trim = 0mm 0mm 0mm 0mm, clip]{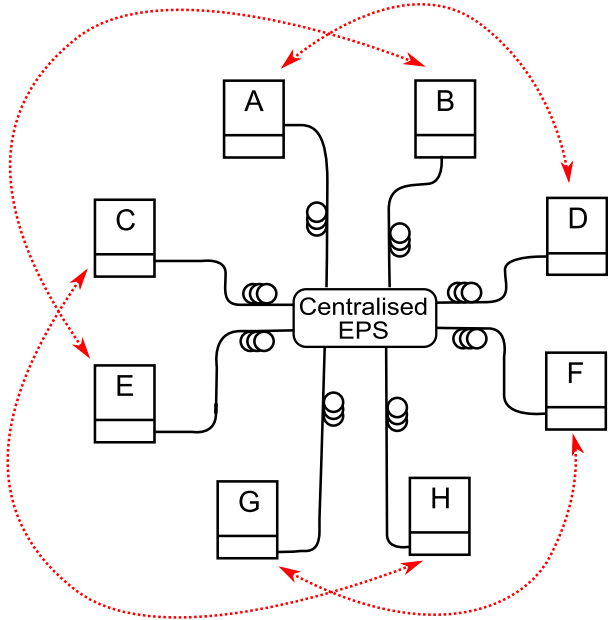}
\caption{Outline of an entanglement distribution network. Each user is connected via a single quantum channel (solid lines) to the centralised entangled-photon-source (EPS). Entanglement (dotted arrows), used for quantum information tasks, can be shared between any two users on request (only certain combinations are shown).}
\label{fig:network}
\end{center}
\end{figure}

We report on the development of an EPS at 1550~nm based on SPDC in lithium-niobate (LiNbO$_3$) waveguides. This approach combines robustness (waveguides and fibre pigtailing) with a clean $\chi^2$ process (low background) and results in a reliable high-brightness source tailored for quantum communication protocols. Moreover, we demonstrate bipartite entanglement distribution (depicted in Figure~\ref{fig:network}), from a centralised EPS in a network for eight users with passive wavelength multiplexing, the highest number reported so far~\cite{kan09, lim08}, and easily extendable to many tens of users by using arrayed-waveguide gratings (AWGs). With a broadband EPS, all users receive a continuous stream of entangled photons at the same time, maximising the use of the fibre infrastructure.  In addition we implement a software-defined network (SDN) for entanglement distribution, where we show active assignment of entanglement to any two users in a 4-user star-like network. We also achieve very high coincidence rates close to 550 counts/s for each entangled channel pair and obtain fidelities of more than 98\% for the entangled states.

In Section~\ref{Setup}, we describe in detail the experimental setup, comprising the source, the implementation of active phase-stabilisation, the components designed for multi-user entanglement distribution, the polarisation analysis as well as the detection modules. We present, in Section~\ref{Results}, the key performance parameters (coincidence rates and conversion efficiency), as well as  the results of tomographic measurements on the entangled channels. With the introduction of optical switches we demonstrate in Section~4 a ``any Alice to any Bob'' 4-user SDN application for entanglement distribution, before concluding in Section~\ref{Concl}.

\section{Setup}
\label{Setup}
\subsection{Source of polarisation entangled photons}
\begin{figure}[t]
\begin{center}
\includegraphics[width=11cm,trim = 0mm 0mm 0mm 0mm,
  clip]{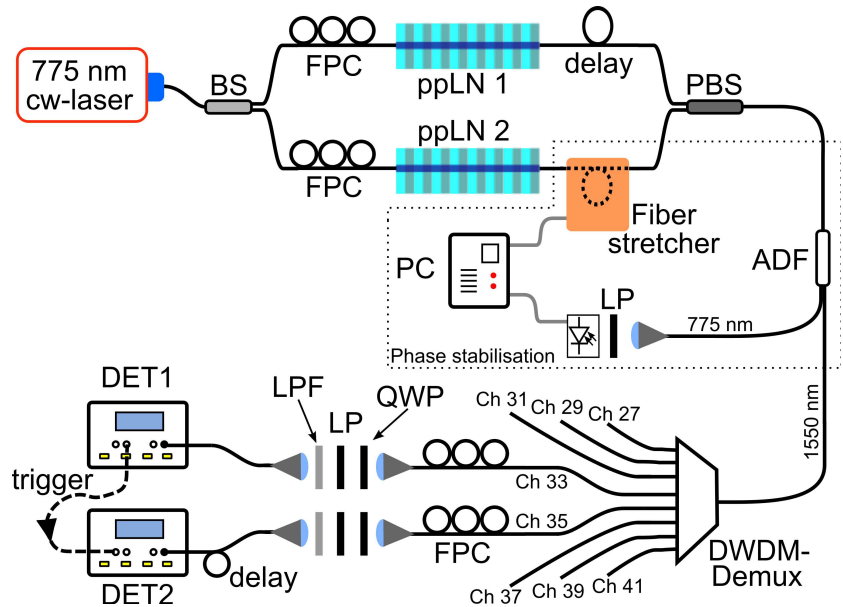}
\caption{Experimental setup with a cw-laser pumping two ppLN crystals in a Mach-Zehnder interferometer to create polarisation-entangled photon pairs. The pairs are split using a DWDM multiplexer, analysed using polarisation optics and detected with InGaAs single photon detectors. The active phase stabilisation of the interferometric part is achieved with the help of a fibre stretcher. }
\label{fig:setup}
\end{center}
\end{figure}

The source, depicted in Figure~\ref{fig:setup}, is based on two 30~mm
long, periodically poled LiNbO$_3$ crystals (ppLN) arranged in a Mach-Zehnder interferometer to yield
polarisation entanglement~\cite{tsu03}. The crystals are type-0 quasi phase-matched (all interacting fields have the same polarisation) to support
SPDC, converting pump photons at 775~nm
to signal and idler photons at 1550~nm, with collinear emission at $\sim60^\circ$C. The phasematching condition at degeneracy is very
broad even for our 30~mm long crystals, leading to a large spectral bandwidth of the
entangled photon pairs of approximately 70~nm.  However, the tight energy uncertainty of the narrowband
pump photons at 775~nm restricts the photons of any pair to be
symmetrically located in frequency around the central wavelength of
1550~nm. Each crystal contains
an inscribed waveguide (proton exchange method) with a $9.7\times7.2$~$\mu$m$^2$ mode field
cross-section at 1550~nm (\emph{HC
  Photonics}), guiding only vertical polarised light. Each
waveguide is fibre-coupled to a $5.4~\mu$m mode field diameter (MFD)
 fibre at the input (single mode for the pump), and a polarisation
maintaining (PM) 1550~nm fibre with a MFD of $10.4~\mu$m at the output (single mode for the generated photon pairs). Both crystals are pumped by a grating-stabilised
narrowband continuous wave (cw) diode laser from \emph{Toptica
  Photonics} (\emph{DL 100}). The laser has a narrow linewidth of
$\sim 1$~MHz, and
can be tuned between $770$~nm and $780$~nm. A fibre port at the laser head
couples around 10~mW of the laser light into a single mode fibre.

 As shown in Figure~\ref{fig:setup}, the pump field is split, at a ratio of $50/50$, into the two spatial
 modes of the interferometer with a fibre-based beam splitter (BS) and is
 then directed to the single mode input fibres of the crystals. Since the quasi phase matching is dependent on the pump polarisation, fibre polarisation controllers (FPC) in each arm are used to adjust the
 polarisation of the incident pump field to generate pairs with vertical polarisation. The horizontal polarisation component of the pump does not contribute to the SPDC process and is also not guided by the waveguide. 
 The PM fibres at the output face of each
 crystal are aligned with their slow axis parallel to the polarisation
 direction of the generated signal/idler pair. Once inside the PM fibre,
 the polarisation of the pair will remain parallel to the slow axis of
 the fibre. The two PM fibres are then combined using a fibre-based
 polarising beam splitter (PBS). This device fuses two input PM fibres
 into a standard single mode fibre, whereby one of the two PM fibres
 is turned by $90^\circ$. Hence, the
 polarisation of the pairs from ppLN~2 is turned by
 $90^\circ$ with respect to the polarisation of the pair from ppLN~1. Since both crystals initially produced pairs with vertical polarisation $\ket{VV}$, the
 entangled state $\ket{\phi} = \ket{HH} + e^{i\theta}\ket{VV}$ is
 created, where $\theta$ is the phase accumulated in the
 interferometer, as discussed in Section 2.2. In our setup, the pump and down-converted photons are
 exclusively transmitted in guided modes (e.g. single-mode fibres and waveguides), making
 realignments obsolete.

To tune the wavelengths of signal and idler photons, the crystal temperature was controlled by electrical heaters. 
Degeneracy, where signal and idler photons have the same spectral properties, was achieved by setting the temperature of ppLN1 and ppLN2 to 66.9 $^\circ$C and 57.5 $^\circ$C respectively, indicating small differences during production.

\subsection{Active phase-stabilisation}
The state of the polarisation entanglement is dependent on the phase difference $\theta$ of both arms of the Mach-Zehnder interferometer. Since this phase is very sensitive to 
changes arising from temperature variations and mechanical vibrations,
an active stabilisation of the phase inside the interferometer was
implemented. A standard telecom add-drop filter (ADF) with a 13~nm wide passband at 1550~nm is used after the
interferometric stage to split the laser light (775~nm) from the
down-converted light ($\sim 1550$~nm). The  pump light is
directed to a phase analyser, consisting of a linear
polariser (LP) and a photo-diode.  Since the pump light passes the whole
Mach-Zehnder interferometer, the change in the phase difference between both arms
is transformed to intensity fluctuations after the polariser (set to
$45\degree$ in the laboratory frame). Phase changes of around $\pi$ per second were found,
clearly indicating the need of an active stabilisation routine. The
output signal of the photodiode was fed into a computer,
which in turn produced voltages to drive a fibre stretcher in
one arm of the interferometer. Stretching the fibre leads to a slight
path length difference and hence a change in the relative phase of the
interferometer. By adding a constant stretching via a software interface, the fibre stretcher is also used to set the phase
$\theta$ of the entangled state on demand. For the rest of this work the phase was fixed to $\theta=0$  to obtain the desired Bell-state:
$\ket{\Phi^{+}} = \frac{1}{\sqrt{2}}\left(\ket{HH} + \ket{VV}\right)$.

\subsection{Multi-user distribution}

The broad bandwidth of  photons generated by the source presented here can be used
to  distribute bipartite entangled states to
many users (8 users in our case) at virtually the same time.  A passive 8-channel demultiplexer (Demux) with channel spacings of 200~GHz matched to the DWDM grid, and
 channel widths ($\Delta\nu$) of 62~GHz (0.5~nm), was used to split the broad
spectrum into 8 output fibres.  By careful tuning the pump laser
wavelength, the central frequency of the down-converted field fell
exactly between the two central channels of the DWDM-Demux. This
configuration resulted in four possible pairs of entangled
channels at wavelengths of 1549.32/1550.92~nm, 1547.72/1552.52~nm,
1546.12/1554.13~nm and 1544.53/1555.75~nm, as shown in the first
column of Table~\ref{tab:channelsresult} and numbered in the standard
ITU channel notation (ch.~27 to ch.~41).  Any such set of two entangled channels can be
directed via single-mode fibres to users wishing to share
entanglement or, as in our setup, to a polarisation analysis and
detection module.
The narrow filtering of the DWDM-Demux also increased the temporal coherence of the single photons, such that dispersion effects in long distance fibre transmission are reduced. The value for polarisation mode dispersion, which leads to depolarisation of the photons, is $\sim 0.4$~ps for a 100~km fibre link~\cite{hub07}, much less than the coherence time of $\sim16$~ps given by the DWDM-Demux bandwidth. For a standard telecom fibre of 100~km length, the chromatic dispersion for such narrow filtering would equal to a temporal photon spread of $\sim 1$~ns, which is also acceptable. Using non-zero dispersion-shifted fibres would narrow the spread even further to below 0.5~ns, the exact value depending on the wavelength of the single photons due to the variation of chromatic dispersion (5.5-10~ps/nm/km) over the C-band (1530~nm to 1565~nm). 
Another advantage in using the DWDM grid is the possibility to run the quantum channel alongside other classical channels in the same fibre, as demonstrated in \cite{Lo10} and hence optimising fibre resources.

\subsection{Polarisation analysis and detection}

To quantify the degree of polarisation entanglement, a full state
tomography was performed for each channel pair. For this measurement,
the photons in each channel passed, in free space, a quarter wave
plate (QWP) followed by a linear polariser (LP) and long-pass filters
(LPF), before being coupled to single mode fibres again. The long-pass
filters were used to remove any residual pump light at
775~nm. Fibre-based polarisation controllers (FPC) were  introduced
before the free space unit to compensate for the unitary rotation of the polarisation state induced by the birefringence of the optical fibre.

The photons of each entangled channel pair were detected by two
single-photon InGaAs avalanche photo diodes (APD), \textit{Id-200}
from \textit{idQuantique}. Since our source was pumped by a cw-laser,
no synchronisation signal, normally used to trigger the InGaAs APDs,
was available. One detector (DET1), with detection efficiency of 10\%
($\eta_{DET1}$), was instead operated with internal triggering
and, upon a detection event, triggered the second detector (DET2),
with detection efficiency of 15\% ($\eta_{DET2}$). The internal triggering
mode on DET1 was achieved by using the longest possible gate of 100~ns
and an internal gate repetition rate ($f_{gate}$) of either 100~kHz or 1~MHz. This implied a
duty cycle ($\eta_{duty}$) of only 1\% or 10\% respectively.  The
photons guided to the second detector were appropriately delayed using
15~m of fibres to coincide with the short 2~ns gate of DET2. The deadtimes of both detectors were set to 
10~$\mu$s.
Free-running avalanche photo diodes~\cite{war09, yan12} or high-efficiency superconduction-detectors~\cite{nam08, nam13} are only now becoming commercially available and were not used in this experiment. Employing these advanced detection schemes would have increased our detected pair rates by 2-3 orders of magnitude compared to the values presented in the next chapter.

\section{Results}
\label{Results}

In contrast to typical SPDC setups where the pump laser is focused onto a small region of the non-liner crystal, the use of a waveguide with strong confinement over a long region of the crystal increases the interaction strength by several orders of magnitude. In the following section we give details of the generation process and detection rates, and characterise entanglement distribution using narrow and broad-band telecom multiplexers. 
\subsection{Coincidence rates and conversion efficiency}
In order to accurately quantify the rates expected from the source,
the total optical loss in the system was characterised using a tunable
laser diode operated around 1550~nm. The coupling efficiency
($\eta_c$) of the waveguide to the fibres (input and output) was
estimated to be $50\pm5\%$, a figure provided by the manufacturer. The
combined transmission of the fibre stretcher and fibre PBS was
measured to be 90\%. For channel 31, the ADF and DWDM-Demux had a combined transmission of 60\%. Lower channel numbers showed
higher transmission, whereas higher channel numbers had lower
transmission. We believe this is due to the inherent geometry of the
DWDM-Demux. Finally, the polarisation analysis
units (QWP, LP and LPF) had a transmission of 75\%. The total
transmission through all optical parts from the
waveguides to the inputs of the detectors was therefore $\eta_{opt}=20\%$.

To characterize the coincidence rates between signal and idler
photons, a single crystal (ppLN 2) was used, and the pump power
($P_{pump}$) was set to $18~\mu$W. The detected single rate (signal
photons) in channel 37 was 2600 counts per seconds (thereafter, c/s)
using DET1 at $1\%$ duty cycle ($f_{gate}=100$~kHz).  As expected from conservation of
energy, idler photons were found in channel 31 and a coincidence rate
($R_c$) of 75~c/s was detected at DET2. The coincidence to single
ratio was found to be $2.9\%$, which compares favourably with the
expected ratio of $3\%$, given by the product of $\eta_{opt}$ and
$\eta_{DET2}$. In order to improve $R_c$ , we increased
the gate frequency of DET1 to 1MHz. This increased the single rate to
38000 c/s (of which 9000 c/s were dark counts) and $R_c$ to 450 c/s, yielding a ratio of $1.5\%$
only. We believe that the single rate is highly increased by
afterpulsing effects. This can also be seen by the nearly 12-fold
increase in the single rate, although only a 6-fold increase would be
anticipated from the actual gate rate on DET1, which was down to
$f_{gate}=620$~kHz due to saturation.  However the coincidence rate did
increase by a factor of 6, which furthermore supports our hypothesis
of afterpulsing. Waveguide ppLN 1 showed a similar behaviour, but with a
lower single and coincidence rate. We attribute this to lossier
fibre-waveguide couplings. To achieve equal coincidence rates between
the two crystals, as required for Bell-state production, the effective
pump power into ppLN 2 was reduced by slightly turning the input
polarisation away from optimum. We also tested the quality of our fibre-pigtailed crystals by second-harmonic-generation (SHG). This reverse process to SPDC converts two pump photons at a wavelength around 1550~nm to a signal photon at 775~nm.  We measured overall SHG-efficiencies of 186\%/W and 218\%/W for the crystals ppLN1 and ppLN2 respectively, again observing a slight difference in the conversion efficiency.

The fibre coupled brightness ($B$) of the source can be calculated by
taking into account the detector efficiencies and bandwidth
($\Delta\nu$) of the photons. Using the data from the $1\%$ duty cycle
run, this results in
$B=R_c/(\eta_{DET1}\,\eta_{DET2}\,\eta_{duty})/P_{pump}/\Delta\nu=4.5\times10^5$~pairs/s/mW/GHz,
which is several orders of magnitude higher than in comparable sources
using nonlinear crystals without waveguides~\cite{hen09,pru12}.

Since the crystals are fibre-coupled with known losses, an attempt was
made to estimate the intrinsic conversion efficiency, i.e. the
probability to create a photon pair given a single pump photon. This
measurement was performed with an input power of 0.62~mW of 775~nm
light, which yielded a SPDC output power of 1.7~nW at 1550~nm as
measured with a standard power meter. The measured conversion
efficiency is hence $2.74\times 10^{-6}$. Including the losses of the
fibre coupling, $\eta_c$, the intrinsic efficiency of the SPDC
inside the waveguide is even larger and estimated to be $1.1\pm 0.1\times 10^{-5}$. Hence, for
every $\sim 10^5$ pump photons, one signal-idler pair is
generated. This is in excellent agreement with a theoretical calculation of the conversion efficiency~\cite{mun07} which yields $1.1\times 10^{-5}$ for the specific ppLN waveguide used in the measurements.

\subsection{Tomographic measurements on entangled channels}

\begin{figure}[t]
\begin{center}
\includegraphics[width=7cm]{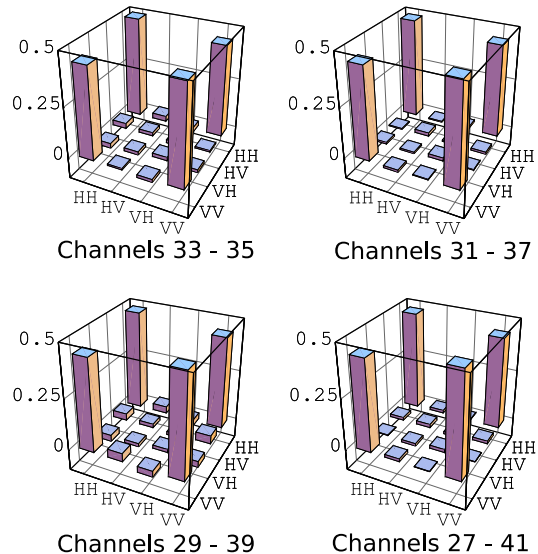}
\caption{Real part of the density matrices obtained from tomography measurements (raw coincidences) for each entangled channel pair. The elements of the imaginary parts are all smaller than 0.07.}
\label{fig:tomo}
\end{center}
\end{figure}

The polarisation state of each entangled channel pair was analysed
using a state tomography measurement~\cite{jam01}. Coincidences were
measured in 16 polarisation settings (HH, HV, HP, HR, VH, VV, VP, VR,
PH, PV, PP, PR, RH, RV, RP and RR), where H,V,P and R stand for
horizontal, vertical, $+45^\circ$ and right circular polarisation
respectively. All coincidences, averaged over 20 seconds, were
recorded with an input power of 18~$\mu$W for each crystal and a duty
cycle of 10\% for DET1.  The maximal coincidence rates measured were
found to be around 450~c/s, as shown in
Table~\ref{tab:channelsresult}. As previously mentioned, each channel
of the DWDM-Demux has different insertion losses affecting the coincidence
rates between channel pairs. Density matrices ($\rho$), calculated
from the raw coincidence rates for each channel pair, can be seen in
Figure~\ref{fig:tomo}. Fidelities ($\bra{\Phi^{+}}\rho\ket{\Phi^{+}}$)
and purities ($\textrm{Tr}(\rho^2)$) for all four entangled channel pairs
were also obtained, and are listed in Table~\ref{tab:channelsresult}. The
fidelities have values around 93\% indicating a high degree of
entanglement of the raw data. This raw fidelity value yields a quantum bit error of $\sim3$\%, low enough to establish a secret key between parties using the BBM92 QKD protocol~\cite{bbm92}.  The observed fidelity is also in very good agreement with a study predicting an entanglement visibility of 93\% for entanglement distribution using a DWDM-Demux~\cite{zaq13}. With substraction of the background of $\sim15$~c/s,
which consists primarily of accidental coincidences from higher order emissions and afterpulsing, 
fidelity and purity values of up to 99\% are observed, also listed in
Table~\ref{tab:channelsresult}. These values show that the source is
producing entangled states at very high fidelity over all eight
channels. Even when considering only one pair of DWDM-Demux channels, our source has the highest reported coincidence rate for a raw fidelity above 90\% in similar $\chi^2$ waveguide sources~\cite{tsu03,tsu05,tom06,lim08a,ino11}.

\begin{table}[t]
\begin{center}
\scalebox{1.2}{%
\begin{tabular}{|c|c|c|c|c|c|}
\hline
Entangled channels   & Coincidence &  & Fidelity without&  & Purity without \\ [-1.2ex]
central wavelengths [nm]&  rate [c/s]&\raisebox{1.5ex}{Fidelity} & background & \raisebox{1.5ex}{Purity} & background \\
\hline
33 - 35   & & & & & \\ [-0.8ex]
1550.92 - 1549.32   & \raisebox{1.5ex}{460}& \raisebox{1.5ex}{0.93} & \raisebox{1.5ex}{0.99} & \raisebox{1.5ex}{0.87} & \raisebox{1.5ex}{0.99} \\
\hline
31 - 37   & & & & &\\ [-0.8ex]
1552.52 - 1547.72    & \raisebox{1.5ex}{450} & \raisebox{1.5ex}{0.92} & \raisebox{1.5ex}{0.98} & \raisebox{1.5ex}{0.89} & \raisebox{1.5ex}{0.98} \\
\hline
29 - 39   & & & & &\\ [-0.8ex]
1554.13 - 1546.12   & \raisebox{1.5ex}{480} & \raisebox{1.5ex}{0.93} & \raisebox{1.5ex}{0.99} & \raisebox{1.5ex}{0.88} & \raisebox{1.5ex}{0.99}\\
\hline
27 - 41   & & & & &\\ [-0.8ex]
1555.75 - 1544.53    & \raisebox{1.5ex}{430} & \raisebox{1.5ex}{0.93} & \raisebox{1.5ex}{0.98} & \raisebox{1.5ex}{0.88} & \raisebox{1.5ex}{0.99}\\
\hline
\end{tabular}}
\caption{Summary of results for the four entangled channel pairs, designated by their ITU numbers and central wavelengths. Coincidence rates and calculated fidelities and purities of the measured states are displayed. Fidelity and purity were calculated both for raw data and after background subtraction.}
\label{tab:channelsresult}
\end{center}
\end{table}
\par

\subsection{Visibility measurements on entangled channels in a CWDM-grid}

The spectrum of our source extends over 5 channels of the widely deployed  CWDM telecom wavelength grid. Less stringent requirements on the wavelength accuracy  leads to cheaper components and high availability paves the way for easy integration of QKD-systems in existing telecom networks. Therefore we replaced the ADF and DWDM-Demux with a 5 channel CWDM-Demux, and sliced the SPDC spectrum into 20~nm wide channels, with a 13~nm passband in each channel. The center channel at 1551~nm was used for the phase stabilisation. In order not to saturate our detectors we reduced the pump power to $1~\mu$W per crystal and recorded a coincidence rate of around 200 c/s which is in agreement with the increased bandwidth of the channels. However visibility measurements for the entangled pairs in the channels with the central wavelength of 1531-1571~nm (1) and 1511-1591~nm (2) showed only values of $V_1=86.8\%$ and $V_2=87.5\%$, respectively. With background subtraction the values increased somewhat to $V_1=91.1\%$ and $V_2=90.9\%$, but were still below the near 100\% mark achieved with the DWDM-Demux.  We believe that the reduction of visibility is caused by the wavelength-dependent birefringence in the fibre. The large bandwidth of the CWDM-Demux causes different wavelenghts to experience different polarisation rotations in the optical fibre. Since our polarisation controller can only correct for a single specific rotation, a large part of the spectrum will deviate from the input polarisation state and hence cause a decrease in visibility. For this reason and because of the effects of chromatic dispersion, discussed above, it is important to optimise the source for a high spectral brightness ($B$)  in order to limit the bandwidth of each channel as much as possible.

\section{Multi-user entanglement distribution --- ``Any Alice to any Bob'' network application}

\begin{figure}[t]
\begin{center}
\includegraphics[width=14cm,trim = 0mm 0mm 0mm 0mm, clip]{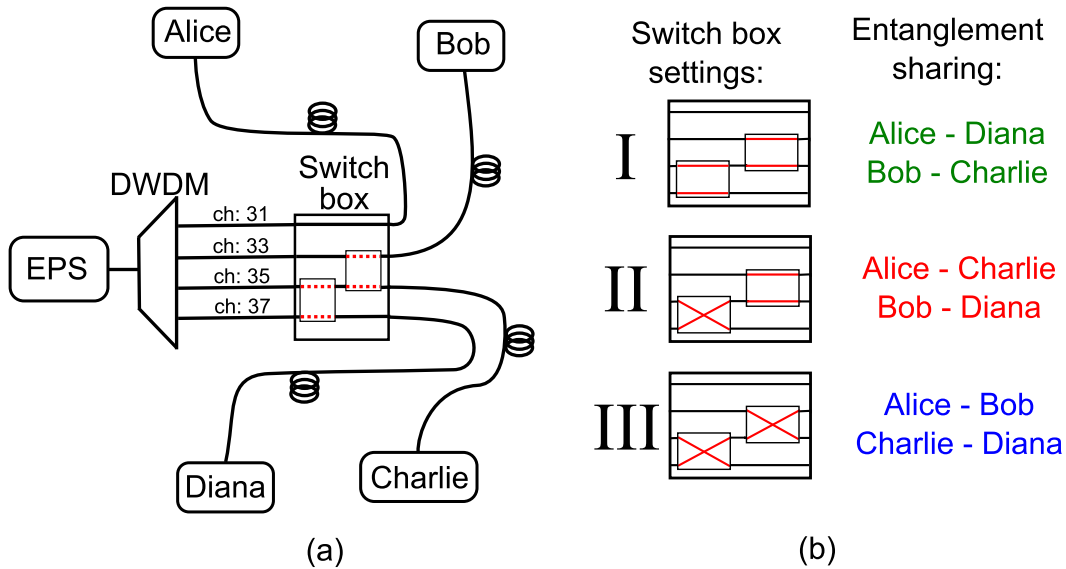}
\caption{Software-defined network (SDN) realisation with two optical switches for active, on-demand distribution of entanglement. Depending on the choice of the switch setting (I, II or III), different user pairs will share entangled photons for quantum information applications.}
\label{fig:switches}
\end{center}
\end{figure}

A single broadband photon source, as presented in the previous
sections, can deliver entangled photon pairs to
many users with the help of a wavelength multiplexer. However this
type of entanglement distribution is static, as the entangled channels
are always connected to the same users.  A software-defined network, where users actively select their partners with whom
they share pairs of entangled photons, can be achieved by adding an
optical switch to the existing setup as shown in Figure~\ref{fig:switches}(a).

We implemented such a SDN for 4 users, by incorporating two
opto-mechanical switches from \emph{Cube Optics} into four of the
fibres leaving the DWDM-Demux. Both switches feature a
$2\times2$ port design (two inputs and two outputs) and are
concatenated to allow for a switching of the entangled inputs to any
output configuration. The distribution of
bipartite entanglement between any two users out of four, was achieved by using three possible switch 
configurations, as schematically depicted in Figure~\ref{fig:switches}(b).

Coincidence rates and polarization correlations for all possible
switch settings were measured again. The rates together with the
calculated state visibilities and purities for all user pairs are
shown in Table~\ref{tab:networkresult}. Coincidence rates were hardly
affected by the addition of the two switches and the visibilities and
purities show again a very high degree of entanglement between any
pair of users.

The number of possible pairs supported by a single source of entangled pairs is only limited by the total emission bandwidth of photons. Using DWDM multiplexers or AWGs with the ITU-recommended 100~GHz spacing, the presented source with its 70~nm overall bandwidth could supply up to N=90 wavelength channels supporting 45 independent pairs of users at the same time. Shorter crystals would lead to even higher number of channels as the SPDC spectrum broadens. The decrease in brightness can easily be compensated by correspondingly higher pumping powers.

The switch box can be expanded to a NxM architecture with N wavelength channels  serving M users by using well-established mathematical methods developed for telecom industry using a combination of 2x2-switches as the basic element. For standard telecommunication architectures, the requirement of non-blocking behaviour is usually important in order to guarantee any possible connectivity from one of the N inputs to any of the M output ports independent of other connections through the switching network. For the purpose of distribution of entangled photon pairs this very restrictive requirement is not necessary, because two users requesting entangled photon pairs to establish quantum correlation, for e.g. a secure QKD-link, are in generally not interested in: (a) which pair of wavelength (i.e. ITU-channels as indicated in Table 1) of the possible N/2 they will receive and (b) which photon of a pair. We therefore believe that entanglement distribution networks scale better in terms of loss than traditional telecommunication networks. In our example shown in Figure 4(b), we distribute N=4 different wavelengths of photons to M=4 users using only two 2x2 switches. A traditional non-blocking Clos-network would require 6 switches (three concatenated switches in parallel), and each photon would therefore experience three times the loss of a single switch.  Moreover, each doubling of the number of input pairs would require the addition of two rows of switches, increasing the overall loss. A constant loss, independent of the number of users, could be achieved using  a mirror array, typically based on Micro-Electro-Mechanical-System (MEMS) technology, which is commercially available for up to N=M=192 channels.

\begin{table}[t]
\begin{center}
\scalebox{1.2}{%
\begin{tabular}{|c|c|c|c|c|c|c|}
\hline

Switch & Entanglement & Coincidence &   & Fidelity without&  & Purity without \\ [-1.2ex]
setting &  sharing & rate [c/s] &\raisebox{1.5ex}{Fidelity} & background & \raisebox{1.5ex}{Purity} & background \\

\hline
 &A-D& 440& 0.93 & 0.98 & 0.89 & 0.98 \\ [-0.8ex]
 \raisebox{1.5ex}{I} & B-C & 480& 0.94 & 0.98 & 0.89 & 0.96 \\
\cline{1-7}
 &A-C& 450& 0.94 & 0.99 & 0.90 & 0.98 \\ [-0.8ex]
 \raisebox{1.5ex}{II}& B-D& 480 & 0.95 & 0.98 & 0.93 & 0.99 \\
\cline{1-7}
 &A-B& 480& 0.92 & 0.98 & 0.86 & 0.96\\ [-0.8ex]
\raisebox{1.5ex}{III}& C-D& 430 & 0.92 & 0.98 & 0.89 & 0.99 \\
\hline
\end{tabular}}
\caption{Summary of results for the three switch settings in the SDN. Each setting results in two 2-party sharing of entangled states. Coincidence rates, calculated fidelities and purities (with and without background counts) for each user pair are listed.}
\label{tab:networkresult}
\end{center}
\end{table}
\par

\par


\section{Conclusion}
\label{Concl}

We demonstrated entanglement distribution by combining an all fibre/waveguide source and standard telecommunication multiplexers connecting eight users in a static configuration.   We also showed dynamically configurable links for entanglement distribution between four users switched by a scalable software-defined network. 
Using the wide bandwidth of degenerate  SPDC, many more users could be fed from a single centralised source
using  DWDM components.  Waveguide technology offers high
SPDC efficiencies and enables the construction of more robust and
integrated optical setups compared to conventional free-space systems
with bulk crystals.  Furthermore, the fibre pigtailing guarantees a
stable coupling of the entangled photons into single mode fibres. As
no alignment is necessary, the source produces a reliable
and autonomous stream of entangled photons with a high coincidence rate.
The narrow channel bandwidth (62~GHz) of the DWDM-Demux also reduces
dispersion effects in optical fibres. Furthermore, the standardisation of the multiplexer allows the incorporation of a quantum channel in existing classical optical networks.

The observed entanglement fidelities after distribution were very high ($93\%$), and reached
up to $99\%$ with background substraction. Although we report the highest coincidence rates for a fibre pigtailed waveguide source, the rate was still limited by InGaAs APDs which were used in a gated mode with a maximal duty cycle of 10\%. However advanced detector technology for the near-infrared is now commercially available  and will remedy this problem in future.

\section{Acknowledgements}
 We would like to thank Michael Hentschel and Sven Ramelow for
technical assistance. This work was supported by the Austrian Science
Foundation FWF (TRP-L135 and SFB-1520), the Natural Sciences and Engineering Research Council of Canada, Canada Foundation
for Innovation, Canadian Institute for Advanced Research, Ontario Research Fund, and Industry Canada. We would also like to thank
the Institute for Quantum Optics and Quantum Information, Vienna, for financial
assistance in obtaining the waveguides. I.H. thanks the Institute for
Quantum Computing, Waterloo, Canada, for hosting her during the
writing-up phase of this article.

\section*{References}

\hbadness=10000
\bibliography{waveguidesource}

\end{document}